\documentclass[aps,pra,superscriptaddress,twocolumn,showpacs]{revtex4}
\usepackage{graphics}
\usepackage{amsmath}
\usepackage{graphicx}
\usepackage{amsfonts}
\usepackage{amssymb}

\begin{document}

\title{Confidential direct communications: a quantum approach using
continuous variables}
\author{Stefano Pirandola}
\affiliation{M.I.T. - Research Laboratory of Electronics, Cambridge MA 02139, USA}
\author{Samuel L. Braunstein}
\affiliation{Computer Science, University of York, York YO10 5DD, United Kingdom}
\author{Seth Lloyd}
\affiliation{M.I.T. - Research Laboratory of Electronics, Cambridge MA 02139, USA}
\affiliation{M.I.T. - Department of Mechanical Engineering, Cambridge MA 02139, USA}
\author{Stefano Mancini}
\affiliation{Dipartimento di Fisica, Universit\`{a} di Camerino, I-62032 Camerino, Italy}
\date{\today }

\begin{abstract}
We consider the problem of privacy in direct communications, showing how
quantum mechanics can be useful to guarantee a certain level of
confidentiality. In particular, we review a continuous variable approach
recently proposed by us [S. Pirandola et al., Europhys. Lett. 84, 20013
(2008)]. Here, we analyze the degree of privacy of this technique against a
broader class of attacks, which includes non-Gaussian eavesdropping.
\end{abstract}

\maketitle

\section{Introduction}

Quantum mechanics provides a nice solution to an old cryptographic problem,
i.e., the key distribution problem \cite{QKD1}. Going further, we consider
whether or not quantum mechanics could profitably be exploited even for
direct confidential communication, without resorting to the use of
pre-distributed private keys. Since many quantum communication protocols
like quantum key distribution (QKD) \cite{QKD1} and quantum teleportation
\cite{Tele} have been extended to continuous variable systems \cite%
{CVTele,CVQKD}, i.e., quantum systems associated to infinite-dimensional
Hilbert spaces \cite{CVbook}, we find it rather natural to address the
problem of direct communication in this framework. Here, an important role
has been played by the bosonic modes of the radiation field, and Gaussian
states. In particular, coherent states of the radiation have become the most
appealing choice for implementing many quantum information tasks. Along this
line, we have shown \cite{QDCepl} how a sender (Alice) can exploit coherent
states of a bosonic mode in order to send confidential messages to a
receiver (Bob), with an acceptable degree of privacy. This is the first
proof-of-principle of a (quasi) confidential quantum direct communication
(QDC) in the framework of continuous variable systems. In particular, this
QDC can be implemented in an easy way, since it exploits the same
\textquotedblleft quantum hardware\textquotedblright\ of the standard
continuous variable QKD, even if this is done via a completely different
logic of classical operations and communications \cite{Phone}. The price one
pays in order to have a simple technique of QDC is that a notion of
\textquotedblleft degree of privacy\textquotedblright\ must replace the one
of unconditional security (used in QKD). This means that we allow a
potential eavesdropper (Eve) to access a limited fraction of the
information, even if this fraction can be evaluated in advance and also made
very small.

The ideal situation for a QDC occurs when Alice and Bob are connected by a
noiseless channel, so that the unique noise they have to correct is due to
the continuous structure of quantum phase-space \cite{Exp}. However, in
general, this is not the case and the honest users must randomly switch
instances of direct communication with instances of statistical checks on
the quantum channel. As soon as they detect the presence of a \textit{%
non-tolerable noise}, they promptly stop the communication. The maximum
noise that can be tolerated is connected to the maximum amount of
information that they are willing to give up to an eavesdropper. In other
words, a good QDC protocol should enable Alice and Bob to communicate the
entire message when the noise is suitably low, while losing a small amount
of information when it is not. According to Ref.~\cite{QDCepl}, the maximum
information that Eve can steal can be made small at will, but at the
expenses of the efficiency of the protocol, corresponding to the ratio of
the number of communicated bits to the number of quantum systems used. An
alternative approach consists in the use of classical error correcting
codes, which makes Eve's perturbation more evident to Alice and Bob's
statistical checks. This approach enables the honest users to reduce the
number of stolen bits while keeping fixed the efficiency of the protocol.
This improvement is proven assuming the model of eavesdropping is also taken
fixed, i.e., Eve is restricted to a Gaussian attack given by a universal
Gaussian cloner.

In the present paper we thoroughly review the results of Ref.~\cite{QDCepl},
giving a more detailed description of the various protocols for QDC,
together with the basic ideas which are behind them. Furthermore, we provide
a deeper analysis of the possible eavesdropping strategies. In particular,
we consider new kinds of attacks which are non-Gaussian and consist in the
intermittent use of Gaussian cloners. These\textit{\ intermittent attacks}
are proven to be more powerful in the eavesdropping of QDC when it is aided
by classical error correction. As a result, the improvement given by the
classical codes is no longer clear if Eve is also allowed to optimize her
strategy. Despite this open problem, the new concepts and the basic schemes
for QDC have still great potentialities to be explored.

The paper is organized as follows. In Sec.~\ref{BasicSection} we review the
basic protocol for QDC together with its Gaussian eavesdropping. In the
following Sec.~\ref{QDCviaCODES}, we review QDC with repetition codes. Its
security analysis is performed in Sec.~\ref{GaussFORcodes} for Gaussian
eavesdropping, and Sec.~\ref{IntermitSEC} for a non-Gaussian generalization.
Finally, after the conclusions of Sec.~\ref{CONCLUsec}, we have added a
discussion on possible variants for QDC in Appendix~\ref{VariantsAPP}.

\section{Basic protocol for quantum direct communication\label{BasicSection}}

\subsection{Continuous variables of a bosonic mode}

Let us consider a bosonic mode with Hilbert space $\mathcal{H}$ and ladder
operators $\hat{a},\hat{a}^{\dagger }$ satisfying $[\hat{a},\hat{a}^{\dagger
}]=1$. Equivalently, this system can be described by a pair of quadrature
operators

\begin{equation}
\hat{q}=\frac{\hat{a}+\hat{a}^{\dagger }}{\sqrt{2}}~,~\hat{p}=\frac{\hat{a}-%
\hat{a}^{\dagger }}{i\sqrt{2}}~,  \label{Quadr}
\end{equation}%
satisfying the dimensionless canonical commutation relation (CCR) $[\hat{q},%
\hat{p}]=i$. From the previous CCR we see that an arbitrary state of the
system $\rho $ must fulfill the uncertainty principle
\begin{equation}
V(\hat{q})V(\hat{p})\geq 1/4~,  \label{HeisPrinc}
\end{equation}%
where $V(\hat{x}):=\mathrm{Tr}(\rho \hat{x}^{2})-[\mathrm{Tr}(\rho \hat{x}%
)]^{2}$ denotes the variance of an arbitrary quadrature $\hat{x}=\hat{q}$ or
$\hat{p}$. In particular, an arbitrary coherent state $|\bar{\alpha}\rangle $
saturates Eq.~(\ref{HeisPrinc}) symmetrically, i.e., $V(\hat{q})=V(\hat{p}%
):=\Delta =1/2$, where the value $1/2$ quantifies the so-called quantum
shot-noise. This is the fundamental noise that affects the \emph{disjoint }%
measurements of the quadratures $\hat{q}$ and $\hat{p}$ of a coherent state
(via homodyne detection \cite{QOptics}). Such a noise is instead doubled to $%
\Delta =1$ when the two quadratures are measured \emph{jointly} (via
heterodyne detection \cite{QOptics}).

According to the Wigner representation, an arbitrary density operator $\rho $
is equivalent to a characteristic function $\chi (\lambda ):=\mathrm{Tr}%
[\rho \hat{D}(\lambda )]$, where $\hat{D}(\lambda ):=\exp (\lambda \hat{a}%
^{\dagger }-\lambda ^{\ast }\hat{a})$ is the \textit{displacement operator}.
Equivalently, $\rho $ can be described by a Wigner function, which is a
quasi-probability distribution defined by the Fourier transform%
\begin{equation}
W(\alpha ):=\int\limits_{\mathbb{C}}\frac{d^{2}\lambda }{\pi ^{2}}\exp
\left( \lambda \mathbf{^{\ast }}\alpha \mathbf{-}\lambda \alpha ^{\ast
}\right) \chi (\lambda )~.  \label{Wig_Complex}
\end{equation}%
In Eq.~(\ref{Wig_Complex}), the Cartesian decomposition of the complex
variable $\alpha =(q+ip)/\sqrt{2}$ provides the real eigenvalues $q$ and $p$
of the quadrature operators of Eq.~(\ref{Quadr}). Such variables span the
phase-space $\mathcal{K}=\{q,p\}$ of the system and, therefore, represent
its fundamental \emph{continuous variables}. For a coherent state $|\bar{%
\alpha}\rangle $, the Wigner function takes the form%
\begin{equation}
W_{|\bar{\alpha}\rangle }(\alpha )=\mathcal{G}_{1/2}(\alpha -\bar{\alpha})%
\text{~,}  \label{Wig_Coherent}
\end{equation}%
where%
\begin{equation}
\mathcal{G}_{V}(\alpha -\bar{\alpha}):=\frac{1}{\pi V}\exp \left( -\frac{%
\left\vert \alpha -\bar{\alpha}\right\vert ^{2}}{V}\right)
\label{Complex_Gaussian}
\end{equation}%
is a complex Gaussian function with mean $\bar{\alpha}$ and variance $V$. As
a consequence, the measurement of the arbitrary quadrature $\hat{x}$\
provides outcomes $x$ which are distributed according to the real Gaussian%
\begin{equation}
G_{\Delta }(x-\bar{x})=\frac{1}{\sqrt{2\pi \Delta }}\exp \left[ -\frac{(x-%
\bar{x})^{2}}{2\Delta }\right] ~,  \label{Gaussian_Real}
\end{equation}%
where $\Delta =1/2$ for homodyne detection, while $\Delta =1$ for heterodyne
detection.

\subsection{Phase-space lattice encoding}

Let us discretize the phase-space $\mathcal{K}$ by introducing a square
lattice whose unit cell has size equal to $2\Omega $ (see Fig.~\ref%
{LatticePic}). An arbitrary cell can be addressed by a pair of integer
indices $(u,u^{\prime })$ and its center specified by the coordinates%
\begin{equation}
q_{u}=2\Omega u~,~p_{u^{\prime }}=2\Omega u^{\prime }~,  \label{Center_Cell}
\end{equation}%
or, equivalently, by the complex amplitude
\begin{equation}
\alpha _{uu^{\prime }}=\frac{q_{u}+ip_{u^{\prime }}}{\sqrt{2}}~.
\end{equation}

\begin{figure}[tbph]
\vspace{-0.3cm}
\par
\begin{center}
\includegraphics[width=0.45\textwidth] {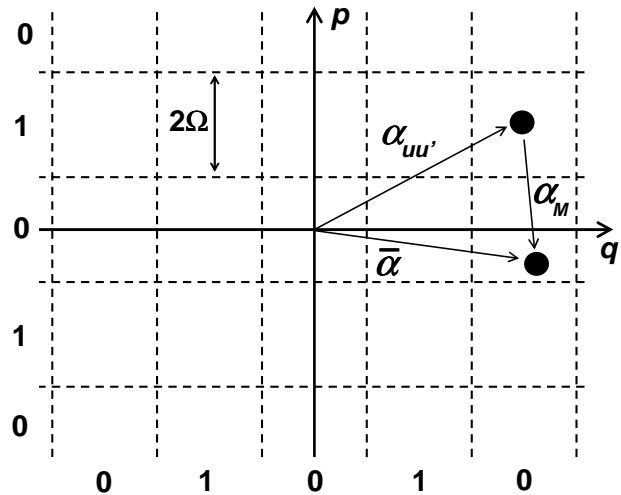}
\end{center}
\par
\vspace{0cm}
\caption{Square lattice of step size $2\Omega $ in the phase-space. The
center of each cell is specified by an amplitude $\protect\alpha %
_{uu^{\prime }}$, where $(u,u^{\prime })$ is a pair of integers representing
the address of the cell. Each address $(u,u^{\prime })$ is associated with a
pair of bits $(U,U^{\prime })$ given by the parities of $u$ and $u^{\prime }$
(see the binary digits at the figure's border). The picture also shows the
masking procedure which adds a mask $\protect\alpha _{M}$ to the amplitude $%
\protect\alpha _{uu^{\prime }}$ in order to create a continuous and Gaussian
signal $\bar{\protect\alpha}$.}
\label{LatticePic}
\end{figure}

Thanks to the introduction of this discrete structure, two bits of
information may be simply encoded in quantum phase-space. In fact, an
arbitrary cell of address $(u,u^{\prime })$ can be associated with a pair of
bits $(U,U^{\prime })$, representing the parities of the indices $u$ and $%
u^{\prime }$. In this approach, Alice encodes two classical bits $%
(U,U^{\prime })$ by choosing a cell whose address $(u,u^{\prime })$ is
randomly selected according to the relations%
\begin{equation}
u=2m+U~,~u^{\prime }=2m^{\prime }+U^{\prime }~,  \label{Random_address}
\end{equation}%
where $m$ and $m^{\prime }$ are random integers \cite{Integers}. Then, she
considers the complex amplitude $\alpha _{uu^{\prime }}$ pointing at the
center of that cell and prepares a corresponding coherent state $|\alpha
_{uu^{\prime }}\rangle $. Such a state is finally sent to Bob, who performs
a heterodyne detection in order to estimate the amplitude $\alpha
_{uu^{\prime }}$ and therefore the encoded information $(U,U^{\prime })$. It
is clear that, even in the presence of a noiseless communication channel,
Bob's decoding cannot be noiseless since the Gaussian shape of the coherent
state spreads over the whole of phase space. Such a spread inevitably leads
to an \emph{intrinsic error} in the decoding process which occurs when the
coherent state is projected by the measurement to wrong peripheral cells.

\begin{figure}[tbph]
\vspace{0cm}
\par
\begin{center}
\includegraphics[width=0.39\textwidth] {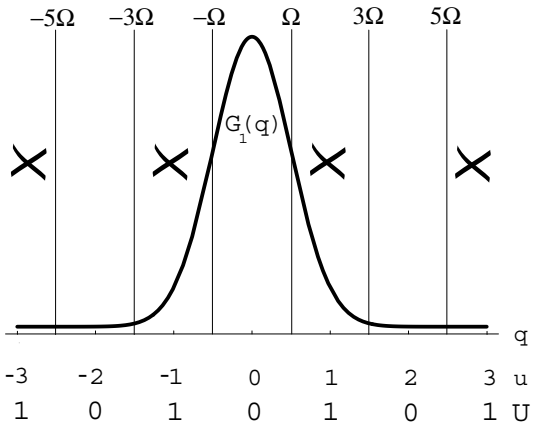}
\end{center}
\par
\vspace{-0.0cm} \caption{Intrinsic error probability
$\protect\varepsilon (\Omega )$ in the decoding of Alice's bit $U$
from the $q$-quadrature of the coherent state. }
\label{GaussErrPic}
\end{figure}

Let us evaluate the probability $\varepsilon $ of an intrinsic error when
Bob decodes Alice's bit $U$ from the position quadrature $\hat{q}$ (the
argument may be repeated for the other quadrature). Since Bob performs a
heterodyne detection on the coherent state, the measured value $q$ will be
distributed around $q_{u}$ according to a Gaussian distribution with
noise-variance equal to $\Delta =1$, i.e., $G_{1}(q-q_{u})$. Suppose, for
simplicity, that $U=0$ is encoded in $q_{u}=0$. According to Fig.~\ref%
{GaussErrPic} an error occurs whenever the measured value $q$ falls in one
of the crossed cells, i.e., having odd index $u=\pm 1,\pm 3,...$ (which
would lead to the incorrect reconstruction of $U=1$ by Bob). Hence, the
probability of an intrinsic error (per quadrature) is equal to
\begin{equation}
\varepsilon (\Omega )=2\sum_{j=0}^{\infty }\,\int_{(4j+1)\Omega
}^{(4j+3)\Omega }dq~G_{1}(q)~.  \label{Intrinsic_error}
\end{equation}%
Now, if we fix a tolerable value for the intrinsic error probability, we
find the corresponding size $\Omega $ to be used for the lattice. In
particular, tolerating $\varepsilon =1\%$ implies adopting $\Omega \simeq
2.57$. On the one hand, the use of a low value for $\varepsilon $\ enables
the honest users to approach noise-free communication. On the other hand, a
large value for $\Omega $ makes the protocol particularly fragile to
eavesdropping. In fact, Eve can optimize her attack on the structure of the
lattice, e.g., by using a non-universal cloner which is optimized on the
centers of the cells. More simply, Eve can detect the state, reconstruct its
cell, and resend another state which is centered in that cell. By resorting
to this intercept-center-resend strategy, Eve is able to remove the noise
most of the time for a sufficiently large $\Omega $. Luckily, we are able to
preclude such strategies by resorting to the classical procedure shown in
the next section.

\subsection{Masking the message and testing the channel}

In order to hide the lattice from Eve, Alice can simply add a \textit{mask}
to her message. After the computation of the \emph{message} amplitude $%
\alpha _{uu^{\prime }}$, Alice classically adds a \emph{mask} amplitude $%
\alpha _{M}$, in such a way that the \emph{total} amplitude $\bar{\alpha}%
:=\alpha _{M}+\alpha _{uu^{\prime }}$ is randomly distributed according to a
complex Gaussian $\mathcal{G}_{V}(\bar{\alpha})$ with large variance $V\gg
\Omega $ (see Fig.~\ref{LatticePic}). Operationally, the whole encoding
procedure goes as follows:

\begin{description}
\item[(1) Lattice Encoding.] Alice encodes the message bits $(U,U^{\prime })$
into a message amplitude $\alpha _{uu^{\prime }}$.

\item[(2) Masking.] Alice picks a signal amplitude $\bar{\alpha}$ from a
wide Gaussian distribution and computes the mask $\alpha _{M}=\bar{\alpha}%
-\alpha _{uu^{\prime }}$ connecting signal and message.

\item[(3) Quantum Preparation.] Alice prepares a signal coherent state $%
\left\vert \bar{\alpha}\right\rangle $ to be sent to Bob.
\end{description}

\begin{figure}[tbph]
\vspace{+0.3cm}
\par
\begin{center}
\includegraphics[width=0.47\textwidth] {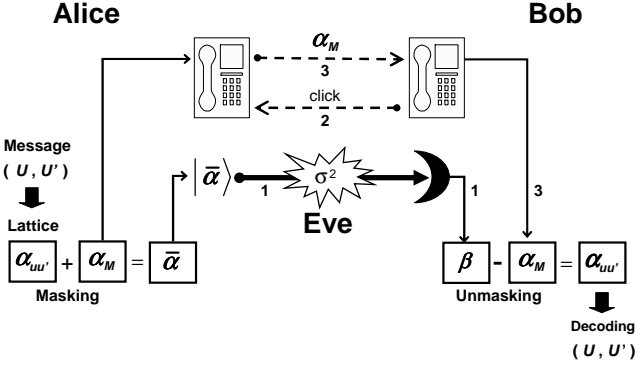}
\end{center}
\par
\vspace{-0.1cm} \caption{\textbf{Message mode (MM).} \ From the
message bits $(U,U^{\prime }) $, Alice computes the message
amplitude $\protect\alpha _{uu^{\prime }}$\ (lattice encoding)\
and then adds the mask $\protect\alpha _{M}$ achieving the signal
amplitude $\bar{\protect\alpha}$. Then, Alice prepares and sends
to Bob the signal state $\left\vert
\bar{\protect\alpha}\right\rangle $, which is heterodyned by Bob
with outcome $\protect\beta $ (step $1$ in the picture). After
detection, Bob classically informs Alice (step $2$) and,
then, Alice classically communicates the mask $\protect\alpha _{M}$ (step $3$%
). Finally, Bob is able to unmask the signal ($\protect\beta -\protect\alpha %
_{M}$), thus reconstructing $\protect\alpha _{uu^{\prime }}$ and, therefore,
$(U,U^{\prime })$.}
\label{MMnoCodesPic}
\end{figure}

\noindent Having prepared the triplet: $\alpha _{uu^{\prime }}$ (message), $%
\alpha _{M}$ (mask) and $\left\vert \bar{\alpha}\right\rangle $ (signal
state), Alice can now perform her quantum and classical communications (see
Fig.~\ref{MMnoCodesPic}). First, Alice sends the signal state $\left\vert
\bar{\alpha}\right\rangle $ to Bob, who heterodynes it with outcome $\beta
\simeq \bar{\alpha}$. Then, after Bob's detection \cite{Bob}, Alice
classically publicizes the mask $\alpha _{M}$. After these two steps, Bob
gets the pair $(\beta ,\alpha _{M})$ from his detection and Alice's
classical communication. Then, Bob is able to \emph{unmask} the signal by
computing $\beta -\alpha _{M}\simeq \bar{\alpha}-\alpha _{M}=\alpha
_{uu^{\prime }}$ and, therefore, estimates the\ message bits $(U,U^{\prime })
$ via lattice decoding.

Clearly, the same decoding steps can be followed by Eve too. However, the
key point is that Eve must choose the probing interaction before knowing the
value of the mask. Since the signal $\bar{\alpha}$ is continuous (Gaussian)
and highly modulated, Eve is prevented from using any kind of interaction
which privileges a particular portion of the phase space. The most natural
choice is therefore a universal Gaussian interaction. A possible model is
given by the universal Gaussian quantum cloning machine (UGQCM) \cite{Cerf}.
Such a machine maps the signal state $\left\vert \bar{\alpha}\right\rangle $
into a pair of output clones $\rho _{B}$ (sent to Bob)\ and $\rho _{E}$
(taken by Eve), each one equal to a Gaussian modulation of $\left\vert \bar{%
\alpha}\right\rangle \left\langle \bar{\alpha}\right\vert $, i.e.,
\begin{equation}
\rho _{K}=\int d^{2}\mu ~\mathcal{G}_{\sigma _{K}^{2}}(\mu )~\hat{D}(\mu )|%
\bar{\alpha}\rangle \langle \bar{\alpha}|\hat{D}^{\dag }(\mu )~,~(K=B,E)~,
\label{Gauss_clon}
\end{equation}%
where the cloning-noise variances $\sigma _{B}^{2}$ and $\sigma _{E}^{2}$
symmetrically affect the quadratures and satisfy the optimality condition%
\begin{equation}
\sigma _{B}^{2}\sigma _{E}^{2}=1/4~,  \label{Cloning_Variances}
\end{equation}%
directly imposed by Eq.~(\ref{HeisPrinc}) \cite{NoteUGQCM}. As a consequence
of Eq.~(\ref{Gauss_clon}), the arbitrary quadrature $\hat{x}$ of the clone $%
K=B,E$ has a marginal distribution equal to $G_{\Delta +\sigma _{K}^{2}}(x-%
\bar{x})$. The performance of the resulting attack will be\ explicitly
studied in the next Section~\ref{GaussAttacksBASIC}.

\begin{figure}[tbph]
\vspace{-0.0cm}
\par
\begin{center}
\includegraphics[width=0.47\textwidth] {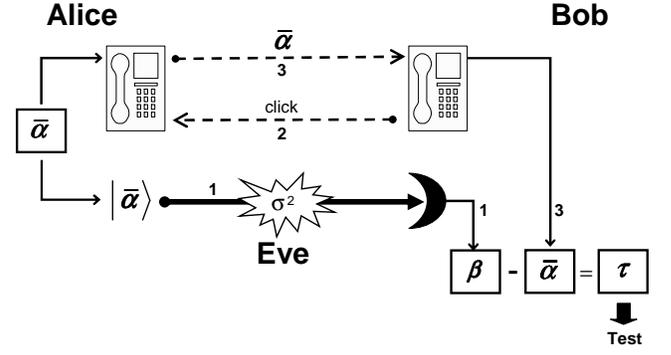}
\end{center}
\par
\vspace{-0.1cm} \caption{\textbf{Control mode (CM).}\emph{\ }Alice
picks up a Gaussian amplitude $\bar{\protect\alpha}$ and prepares
a coherent state $\left\vert \bar{\protect\alpha}\right\rangle $.
Such a state is sent to Bob and heterodyned with outcome
$\protect\beta $ (step 1 in the picture). Then, Bob classically
informs Alice (step 2) and Alice communicates the value of the
signal $\bar{\protect\alpha}$ (step 3). Finally, Bob computes the
test variable $\protect\tau :=\protect\beta -\bar{\protect\alpha}$
to infer the amount of noise $\protect\sigma ^{2}$ in the
channel.} \label{CMnoCodesPic}
\end{figure}

The above procedure of directly communicating message bits is called the
\emph{message mode} (MM) of the protocol. However, Alice and Bob must also
understand how much the channel is perturbed during the communication
process, in order to control the amount of information which is left to a
potential eavesdropper. Assuming an attack with UGQCM, this corresponds to
estimating the value of the noise $\sigma _{B}^{2}:=\sigma ^{2}$ which is
added by Eve to the channel. A real-time check of this noise is possible if
Alice randomly switches from instances of message mode to suitable instances
of \emph{control mode} (CM). In control mode, Alice does not process any
text message but only prepares and sends the signal state $\left\vert \bar{%
\alpha}\right\rangle $ (see Fig.~\ref{CMnoCodesPic}). Then, after Bob's
detection (outcome $\beta $), Alice communicates the value $\bar{\alpha}$ of
the signal amplitude. At that point, Bob extracts from $(\beta ,\bar{\alpha})
$ the actual value of the\emph{\ test variable} $\tau :=\beta -\bar{\alpha}$
which is then used to infer the total noise $\Delta _{B}=1+\sigma ^{2}$
affecting the signal. As soon as they recognize a non-tolerable noise, i.e.,
$\sigma ^{2}>\tilde{\sigma}^{2}$ for some threshold noise $\tilde{\sigma}^{2}
$, they stop the communication. Hereafter, we assume a zero-tolerance
protocol where no added noise is tolerated on the channel, i.e., $\tilde{%
\sigma}^{2}=0$. We shall see that the QDC protocol can be applied in
realistic situations even with such a strict condition \cite{Zero}.

Let us show how the real-time check of the quantum channel works in detail.
Let us consider the Cartesian decomposition $\tau =(q+ip)/\sqrt{2}$ of Bob's
test variable. If the channel is noiseless, then the arbitrary quadrature $%
x=q$ or $p$ is only affected by heterodyne noise $\Delta =1$, i.e., it is
distributed according to a Gaussian distribution $G_{1}(x)$. By contrast, if
Eve perturbs the quantum channel using a UGQCM with noise $\sigma ^{2}\neq 0$%
, then $x$ follows a wider Gaussian distribution $G_{1+\sigma ^{2}}(x)$. By
reconstructing the experimental distribution of $x$ from consecutive
outcomes $\{x_{1},x_{2},\cdots \}$, Bob must therefore distinguish between
the two theoretical distributions $G_{1}(x)$ and $G_{1+\sigma ^{2}}(x)$. In
other words, Bob must distinguish between the two hypotheses%
\begin{equation}
\left\{
\begin{array}{c}
H_{0}:~\mathrm{(Eve=no)~}\Leftrightarrow \sigma ^{2}=0~, \\
H_{1}:~\mathrm{(Eve=yes)}\Leftrightarrow \sigma ^{2}\neq 0~.%
\end{array}%
\right.  \label{Hp_Test}
\end{equation}%
Let us fix the confidence level\ $r$ of this hypothesis test, i.e., the
probability to reject $H_{0}$ though it is true. This level must be
sufficiently low (e.g., $r=5\times 10^{-7}$), so that the direct
communication can be effectively completed in absence of Eve. For each
instance of control mode, Bob makes two independent tests, one for each
quadrature. Hence, after $M$ control modes, he has collected $2M$
quadratures values $\{q_{1},p_{1},\cdots
,q_{M},p_{M}\}:=\{x_{1},x_{2},\cdots ,x_{2M-1},x_{2M}\}$ and he can
construct the estimator
\begin{equation}
v:=\sum_{l=1}^{2M}x_{l}^{2}~.  \label{Estimator}
\end{equation}%
Then, the hypothesis $H_{0}$ is accepted if and only if%
\begin{equation}
v<\mathcal{V}_{2M,1-r}~,  \label{AcceptsH0}
\end{equation}%
where $\mathcal{V}_{i,j}$ is the $j$th quantile of the $\chi ^{2}$
distribution with $i$ degrees of freedom. In other words, Alice and Bob
continue their direct communication in MM as long as the condition of Eq.~(%
\ref{AcceptsH0}) is satisfied in CM.

\subsection{Gaussian eavesdropping\label{GaussAttacksBASIC}}

\begin{figure}[tbph]
\vspace{+0.4cm}
\par
\begin{center}
\includegraphics[width=0.47\textwidth] {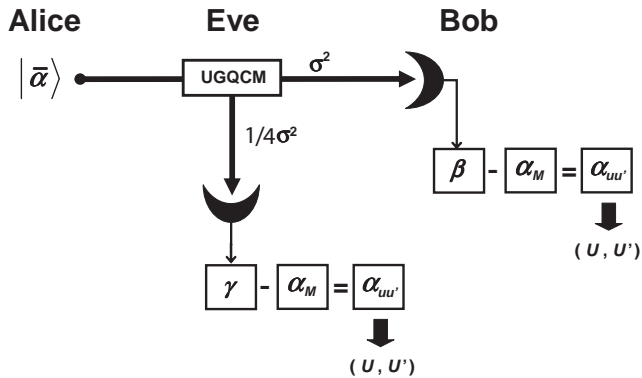}
\end{center}
\par
\vspace{-0.1cm} \caption{Individual UGQCM attack. Eve uses a UGQCM
to eavesdrop the quantum
communication line. Eve heterodynes her clone to get her estimate $\protect%
\gamma $ of the signal amplitude $\bar{\protect\alpha}$. After the public
unmasking of the signal, Eve estimates $\protect\alpha _{uu^{\prime }}$ and,
therefore, the message bits $(U,U^{\prime })$.}
\label{UGQCMattackPic}
\end{figure}

Let us explicitly analyze what happens when the quantum communication
channel is subject to Gaussian eavesdropping via a UGQCM. In an individual
UGQCM attack (see Fig.~\ref{UGQCMattackPic}), Eve clones the signal input
and, then, heterodynes her output to derive her estimate $\gamma $ of the
signal amplitude $\bar{\alpha}$. After the release of the mask's value $%
\alpha _{M}$, Eve infers the message amplitude $\alpha _{uu^{\prime }}$ and,
therefore, the input bits $(U,U^{\prime })$. In this process, Eve introduces
an added noise $\sigma ^{2}$ on the Alice-Bob channel (i.e., $\Delta
_{B}=1+\sigma ^{2}$), while her output is affected by a total noise equal to
$\Delta _{E}=1+(4\sigma ^{2})^{-1}$. This is the sum of the cloning noise $%
\sigma _{E}^{2}=(4\sigma ^{2})^{-1}$, given by the UGQCM, and the
measurement noise $\Delta =1$, given by the heterodyne detector.

First of all, we must evaluate the probability of accepting $H_{0}$ (hence
continuing the communication) notwithstanding the presence of Eve. In other
words, we must compute the probability $\Pi _{M}(\sigma ^{2})$ that Eve
evades $M$ control modes while introducing a noise $\sigma ^{2}\neq 0$.
After $M$ control modes, the estimator of Eq.~(\ref{Estimator}) follows the
distribution%
\begin{equation}
P_{M}(v)=\frac{v^{M-1}}{2^{M}(M-1)!(1+\sigma ^{2})^{M}}\exp \left[ -\frac{v}{%
2(1+\sigma ^{2})}\right] ~.
\end{equation}%
As a consequence, the probability to accept $H_{0}$ is equal to%
\begin{eqnarray}
\Pi _{M}(\sigma ^{2}) &=&\int_{0}^{\mathcal{V}_{2M,1-r}}d\nu ~P_{M}(v)
\notag \\
&=&\frac{\Gamma (M,0)-\Gamma \left( M,\frac{\mathcal{V}_{2M,1-r}}{2(1+\sigma
^{2})}\right) }{(M-1)!}~,  \label{Survival_Probability}
\end{eqnarray}%
where
\begin{equation}
\Gamma (z,a):=\int_{a}^{+\infty }dt~t^{z-1}e^{-t}
\end{equation}%
is the incomplete gamma function.

\begin{figure}[tbph]
\vspace{0cm}
\par
\begin{center}
\includegraphics[width=0.48\textwidth] {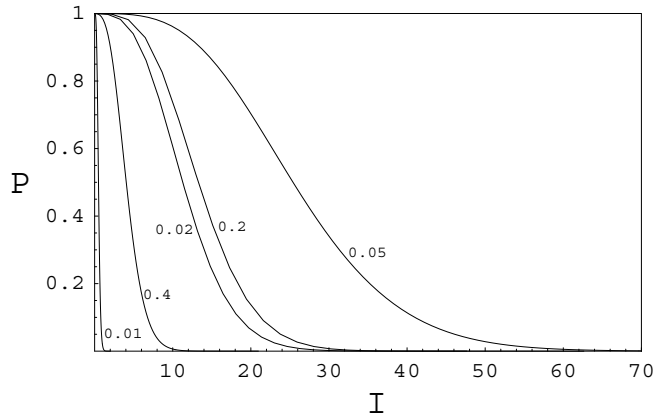}
\end{center}
\par
\vspace{-0.4cm}
\caption{Survival probability $P$ versus the number of stolen bits $I$. QDC\
with parameters $\Omega =2.57$ and $c=69/70$ (so that $\protect\varepsilon %
=1\%$ and $\mathcal{E}=1/35$). The curves refer to individual UGQCM attacks
with different values of added noise $\protect\sigma ^{2}$.}
\label{NoCodesPic}
\end{figure}

Besides Eve's survival probability of Eq.~(\ref{Survival_Probability}), we
must also evaluate the amount of information that Eve can get during her
undetected life on the channel. Such a quantity is limited by the\emph{\
total noise} experienced by Eve, which is equal to $\Delta _{E}=1+(4\sigma
^{2})^{-1}$. For a given $\Delta _{E}$, we now calculate the average
information Eve can steal in a single run of MM. Starting from the outcome
of the measurement $\gamma $ and the knowledge of the mask $\alpha _{M}$,
Eve estimates Alice's amplitude $\alpha _{uu^{\prime }}$ via the variable $%
\gamma -\alpha _{M}$. The corresponding quadrature $x$ will be distributed
according to a Gaussian distribution $G_{\Delta _{E}}(x-x_{u})$. Then, by
repeating the same derivation leading to Eq.~(\ref{Intrinsic_error}), we can
compute Eve's error probability in decoding Alice's bit ($U$ or $U^{\prime }$%
), which is equal to%
\begin{equation}
p(\Delta _{E})=2\sum_{j=0}^{\infty }\int_{(4j+1)\Omega }^{(4j+3)\Omega
}dx~G_{\Delta _{E}}(x)~.  \label{Eve_error_probability}
\end{equation}%
Let us assume that every message bit is a bit of information, i.e., the
input message is not compressible. As a consequence, the average amount of
information which is eavesdropped in a single MM is given by%
\begin{equation}
I_{AE}(\Delta _{E})=2\{1-H[p(\Delta _{E})]\}~,  \label{Stolen_Info}
\end{equation}%
where
\begin{equation}
H(p):=-p\log p-(1-p)\log (1-p)~.  \label{Shannon_Binary}
\end{equation}%
By replacing $\Delta _{E}=1+(4\sigma ^{2})^{-1}:=\Delta _{E}(\sigma ^{2})$
in Eq.~(\ref{Stolen_Info}), we derive $I_{AE}=I_{AE}(\sigma ^{2})$, i.e.,
the average amount of information which is stolen for a given noise $\sigma
^{2}$ in Alice-Bob channel. Such a quantity can be directly combined with $%
\Pi _{M}=\Pi _{M}(\sigma ^{2})$ of Eq.~(\ref{Survival_Probability}). This
means that we can express Eve's survival probability as a function of the
stolen information. In fact, let us fix the probability $c$ of a control
mode, so that $N$ runs of the protocol can be divided into $cN$ control
modes and $(1-c)N$ message modes, on average. As a consequence, Eve's
survival probability is equal to
\begin{equation}
\Pi _{cN}(\sigma ^{2}):=P~,  \label{Eve_surv_prob}
\end{equation}%
and the average number of stolen bits is equal to
\begin{equation}
(1-c)NI_{AE}(\sigma ^{2}):=I~.  \label{Eve_Stolen_Info}
\end{equation}%
Then, for every $\sigma ^{2}$, we can consider the function $P=P(I)$. In
particular, let us fix $c=69/70$, so that the protocol has efficiency%
\begin{equation}
\mathcal{E}:=\frac{\text{number of bits}}{\text{number of transmitted systems%
}}=\frac{1}{35}~.
\end{equation}%
For several values of $\sigma ^{2}$, we can (numerically) evaluate the
function $P=P(I)$ as shown in Fig.~\ref{NoCodesPic}. From this figure, we
can see that, if the noise is low, e.g., $\sigma ^{2}=0.01$, Eve steals very
little information ($\simeq 1$ bit) while Alice and Bob complete an almost
noiseless QDC. In particular, Alice is able to transmit $\simeq 1.5\times
10^{4}$ bits of information by using $N\simeq 5\times 10^{5}$ systems.
Notice that the maximum length of the QDC is roughly bounded by the
verification of $r^{-1}$ hypothesis tests and, therefore, it is limited to
about $4(1-c)(cr)^{-1}$ bits (i.e., $\simeq 1.2\times 10^{5}$ bits or $%
\simeq 4\times 10^{6}$ systems using the above parameters). If the attack is
more noisy (e.g., $\sigma ^{2}=1$), Eve again steals little information ($%
\simeq 1$ bit). In such a case, in fact, Eve is promptly detected by the
honest parties who, however, are prevented from exchanging information
(denial of service). According to Fig.~\ref{NoCodesPic}, Eve's best strategy
corresponds to using a UGQCM\ with $\sigma ^{2}\simeq 1/20$, so that she can
steal a maximal amount of about $80$ bits before being revealed (using a cut
off of $P=1\%$). In such a case, Alice transmits $\simeq 630$ bits by using $%
N\simeq 2.2\times 10^{4}$ systems.

How can we decrease the maximal amount of stolen information? The simplest
solution consists in increasing the control mode probability $c$, so that
the possible presence of Eve is detected before sending too many bits.
Clearly, this approach has a price to pay, which is a decrease of the
efficiency $\mathcal{E}$ of the protocol. An alternative solution consists
of making the decoding more sensitive to the presence of added noise. Such
an approach is possible by introducing classical error correcting codes, and
its pros and cons are explored in the following section. In particular, this
solution is good against Gaussian attacks but its advantages are not
completely clear in the presence of non-Gaussian attacks.

\section{Quantum direct communication with repetition codes\label%
{QDCviaCODES}}

\subsection{The basic idea in using classical codes\label{IdeaCodes}}

In the basic scheme of QDC with continuous variables, a noiseless
communication is possible up to an intrinsic error probability $\varepsilon $
which depends on the step $\Omega $ of the phase-space lattice. In
particular, such a probability decreases for increasing $\Omega $. An
alternative way for decreasing $\varepsilon $ consists of leaving $\Omega $
unchanged while introducing a classical error correcting code for
encoding/decoding. Such procedures are equivalent for a noiseless channel,
since $\varepsilon $ is sufficiently small and the codes work very well in
that case. However, the scenario is different as the channel becomes nosier.
In such a case, in fact, the correcting codes have a non-linear behavior
which makes their performance rapidly deteriorate. Such a non-linear effect
can be exploited to critically split the correction capabilities, and
therefore the information gains, between Alice-Bob channel and Alice-Eve
channel.

\begin{figure}[tbph]
\vspace{+0.2cm}
\par
\begin{center}
\includegraphics[width=0.48\textwidth] {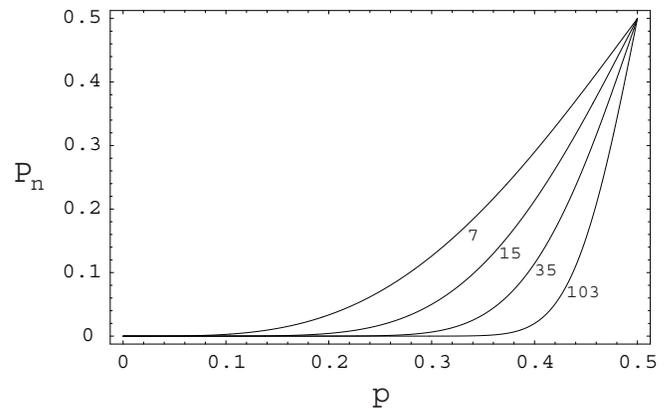}
\end{center}
\par
\vspace{-0.5cm}
\caption{Probability of an uncorrectable error $P_{n}$ versus the single
bit-flip probability $p$. Here, we consider repetition codes with $%
n=7,15,35,103$.}
\label{repetitionPic}
\end{figure}

Let us consider the simple case of an $n$-bit repetition code, where an
input bit $U=\{0,1\}$ is encoded into a logical bit $\bar{U}=\{\bar{0},\bar{1%
}\}$ of $n$ physical bits via the codewords%
\begin{equation}
\bar{0}=\underset{n}{\underbrace{00\cdots 0}}~,~\bar{1}=\underset{n}{%
\underbrace{11\cdots 1}}~.
\end{equation}%
By choosing an odd $n=2m+1$ (with $m=1,2,\cdots $), we can apply a
non-ambiguous majority voting criterion. This means that every bit-flip
error of weight $t<m+1$ is correctable, while every bit-flip error of weight
$t\geq m+1$ is not. Let us now consider a memoryless channel, where each
physical bit is perturbed independently with the same bit-flip probability $p
$, as happens in the case of individual Gaussian attacks. Then, the
probability of an uncorrectable error is simply given by%
\begin{equation}
P_{n}(p)=\sum_{k=m+1}^{n}\left(
\begin{array}{c}
n \\
k%
\end{array}%
\right) p^{k}(1-p)^{n-k}~.  \label{uncorregible_nbits}
\end{equation}%
As it is evident from Fig.~\ref{repetitionPic}, the correction capability of
the $n$-bit repetition code rapidly worsens as the single bit-flip
probability approaches $1/2$. This is due to the non-linear behavior of $%
P_{n}=P_{n}(p)$ which becomes more manifest when $n$ increases. In
particular, for a sufficiently large $n$, the curve displays a critical
point $\tilde{p}$ after which the correction capability suddenly starts to
deteriorate very quickly (e.g., $\tilde{p}\simeq 0.3$ for $n=35$ and $\tilde{%
p}\simeq 0.4$ for $n=103$). Exactly these critical points can be exploited
to improve the QDC, by transforming the communication protocol into a
threshold process, where the sensitivity to added noise is remarkably
amplified.

For a repetition code of fixed length $n$, we have a corresponding critical
value $\tilde{p}$. Then, we can choose a lattice whose step is critical.
This is the value $\tilde{\Omega}$ such that the intrinsic error probability
is critical, i.e., $\varepsilon (\tilde{\Omega})=\tilde{p}$. On the one
hand, when the channel is noiseless, Bob is able to recover the codewords
and reconstruct the logical bit with a very low error probability $%
P_{B}=P_{n}(\tilde{p})$. On the other hand, when the channel is noisy,
Alice's information is split into two sub-channels: the Alice-Bob channel,
with added noise $\sigma _{B}^{2}:=\sigma ^{2}$, and the Alice-Eve channel,
with added noise $\sigma _{E}^{2}=(4\sigma ^{2})^{-1}$. The corresponding
error probabilities are respectively given by%
\begin{equation}
P_{B}=P_{n}(\tilde{p}+p_{B})~,~P_{E}=P_{n}(\tilde{p}+p_{E})~,
\label{Logic_err_prob}
\end{equation}%
where $p_{B}=p_{B}(\sigma _{B}^{2})$ and $p_{E}=p_{E}(\sigma _{E}^{2})$ are
monotonic functions of the added noises (and, therefore, linked by the
uncertainty principle). Now, if Eve tries to hide herself by perturbing the
Alice-Bob channel with a relatively small $p_{B}$, then her dual $p_{E}$
will always be big enough to perturb $\tilde{p}$ into the nonlinear region.
As a consequence, Eve will tend to experience $P_{E}\simeq 1/2$ gaining her
negligible information.

\begin{figure}[tbph]
\vspace{0cm}
\par
\begin{center}
\includegraphics[width=0.35\textwidth] {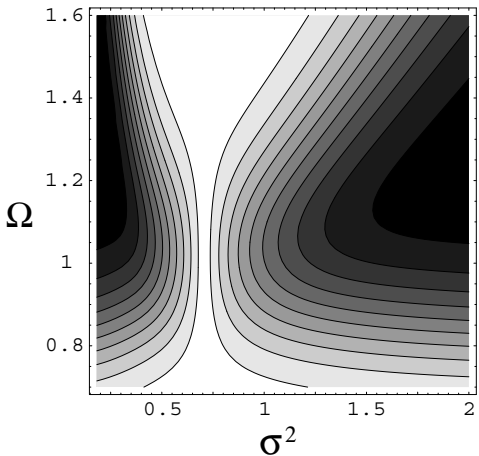}
\end{center}
\par
\vspace{-0.2cm}
\caption{Difference $D:=\left\vert I_{AB}-I_{AE}\right\vert $ on the plane $%
(\Omega ,\protect\sigma ^{2})$. The values increase from $D=0$
(white area) to $D=1$ (black area). Notice how the areas of
separation (black areas) are broader at the critical value
$\tilde{\Omega}\simeq 1$.} \label{ExclusivePlotPic}
\end{figure}

Let us explain the previous point in terms of mutual information. In
particular, let us fix the repetition code to the value $n=35$, so that we
have $\tilde{p}\simeq 0.3$ and a corresponding critical value $\tilde{\Omega}%
\simeq 1$ for the lattice. Starting from an arbitrary $\Omega $, one can see
that $\tilde{\Omega}$ is indeed optimal for Alice and Bob. For every bit of
information which is encoded by Alice, the amount of information decoded by
Bob and Eve is respectively given by
\begin{eqnarray}
I_{AB} &=&1-H(P_{B}):=I_{AB}(\Omega ,\sigma ^{2})~, \\
I_{AE} &=&1-H(P_{E}):=I_{AE}(\Omega ,\sigma ^{2})~,  \label{Eve_Iae_Codes}
\end{eqnarray}%
where $P_{B}$\ and $P_{E}$ are the logical error probabilities in Eq.~(\ref%
{Logic_err_prob}) with $n=35$. Since the added noises satisfy the
uncertainty relation of Eq.~(\ref{Cloning_Variances}), a similar relation
holds for the mutual informations, i.e.,
\begin{equation}
I_{AB}(\Omega ,\sigma ^{2})+I_{AE}(\Omega ,\sigma ^{2})=\mu (\Omega ,\sigma
^{2})~,
\end{equation}%
where $\mu (\Omega ,\sigma ^{2})\leq 2$ is numerically shown in Fig.~\ref%
{SumInfosPic}. Let us also consider the difference of information%
\begin{equation}
D(\Omega ,\sigma ^{2}):=\left\vert I_{AB}(\Omega ,\sigma ^{2})-I_{AE}(\Omega
,\sigma ^{2})\right\vert ~.
\end{equation}%
Such a quantity is a point-by-point measure of how much $I_{AB}$ and $I_{AE}$
are different. In particular, the maximum value $D=1$ corresponds to the
maximal separation $\{I_{AB},I_{AE}\}=\{0,1\}$ or $\{1,0\}$. As we can see
from Fig.~\ref{ExclusivePlotPic}, the points $(\Omega ,\sigma ^{2})$ with $%
\Omega =1$ (i.e., with $\Omega \simeq \tilde{\Omega}$) corresponds to the
broadest areas of separation. In other words, the critical condition $\Omega
\simeq \tilde{\Omega}$ enhances the split between $I_{AB}$ and $I_{AE}$.
\begin{figure}[tbph]
\vspace{+0.5cm}
\par
\begin{center}
\includegraphics[width=0.35\textwidth] {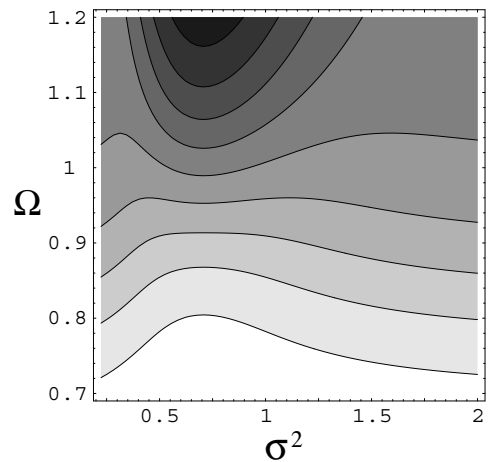}
\end{center}
\par
\vspace{-0.2cm}
\caption{Sum of the mutual informations $I_{AB}+I_{AE}$ on the plane $%
(\Omega ,\protect\sigma ^{2})$. The values increase from 0 (white
area) to 2 (black area). Note how the behavior of the borders
changes around the critical value $\tilde{\Omega}\simeq 1$.}
\label{SumInfosPic}
\end{figure}

\subsection{Protocol with repetition codes\label{sQDCcodedSection}}

Let us explicitly show how to use an $n$-bit repetition code for
encoding/decoding. This is possible by simply adding pre-encoding and
post-decoding classical steps to the basic protocol of Sec.~\ref%
{BasicSection}. The message bits $(U,U^{\prime })$ are pre-encoded into a
pair of logical bits%
\begin{equation}
\bar{U}=U_{1}U_{2}\cdots U_{n}~,~\bar{U}^{\prime }=U_{1}^{\prime
}U_{2}^{\prime }\cdots U_{n}^{\prime }~,
\end{equation}%
via the $n$-bit repetition code. Each pair of physical bits $%
(U_{k},U_{k}^{\prime })$ is then subject to the same encoding as before,
i.e., lattice encoding $(U_{k},U_{k}^{\prime })\rightarrow \alpha
_{u_{k}u_{k}^{\prime }}:=\alpha _{k}$, masking $\alpha _{k}\rightarrow
\alpha _{k}+\alpha _{M}=\bar{\alpha}$ and quantum preparation $\bar{\alpha}%
\rightarrow \left\vert \bar{\alpha}\right\rangle $. Then, after $n$ message
modes, Bob will have collected perturbed versions of the $n$ pairs $%
(U_{1},U_{1}^{\prime }),\cdots ,(U_{n},U_{n}^{\prime })$. By applying
standard error recovery (majority voting), he will then perform the
post-decoding of $(U,U^{\prime })$. In the same way as before, these
instances of message mode (each one carrying a single physical bit of a
codeword) must be randomly switched with instances of control mode, where
Alice skips encoding and simply sends Gaussian signals $\bar{\alpha}$ for
testing the channel (exactly as in Fig.~\ref{CMnoCodesPic}).

Let us choose a repetition code with $n=35$, and a lattice with $\Omega
=1\simeq \tilde{\Omega}$. The latter choice implies an intrinsic error
probability $\varepsilon $, in decoding the physical bits $%
(U_{k},U_{k}^{\prime })$, which is equal to the critical value of the code $%
\tilde{p}\simeq 32\%$. After error recovery, the intrinsic error probability
$\bar{\varepsilon}$ affecting the logical bits $(\bar{U},\bar{U}^{\prime })$
is sufficiently low and corresponds to $P_{35}(\tilde{p})\simeq 1\%$. Then,
let us also choose $c=1/2$ for the control mode's probability, so that we
have an efficiency $\mathcal{E}=1/35$. Notice that the values of $\bar{%
\varepsilon}$ and $\mathcal{E}$ correspond to the ones chosen for the basic
protocol of Sec.~\ref{BasicSection} (where $\bar{\varepsilon}=\varepsilon $
of course). Such parameters equalize the performances of the two protocols
in the case of noiseless quantum channel. As a consequence, we are in a
situation to make a fair comparison between the protocols when malicious
noise is present on the channel.

\subsection{Gaussian eavesdropping\label{GaussFORcodes}}

Let us analyze the effect of an individual UGQCM attack. On every cloned
system, affected by a noise $\sigma _{E}^{2}=(4\sigma ^{2})^{-1}$, Eve
detects the complex amplitude $\gamma $ via heterodyne detection. Then, she
estimates the signal amplitude $\bar{\alpha}$ up to a total noise $\Delta
_{E}=1+\sigma _{E}^{2}$. After Alice's declaration of the mask $\alpha _{M}$%
, Eve derives the message amplitude and, therefore, a pair of physical bits $%
(U_{k},U_{k}^{\prime })$. Each physical bit will be affected by an error
probability $p(\Delta _{E})$ as in Eq.~(\ref{Eve_error_probability}). After $%
n$ eavesdropped message modes, Eve will be able to decode Alice's logical
bits $(\bar{U},\bar{U}^{\prime })$ by majority voting, up to an error
probability $P_{E}=P_{n}[p(\Delta _{E})]$ [see Eq.~(\ref{uncorregible_nbits}%
)]. For each logical bit, the acquired information is simply equal to $%
1-H(P_{E})$. As a consequence, for each message mode, Eve acquires on average%
\begin{equation}
I_{AE}(\sigma ^{2})=2[1-H(P_{E})]/n  \label{Eve_stolen_codes}
\end{equation}%
bits of information (simply because $2$ logical bits are sent via $n$
physical systems).

\begin{figure}[tbph]
\vspace{+0.3cm}
\par
\begin{center}
\includegraphics[width=0.47\textwidth] {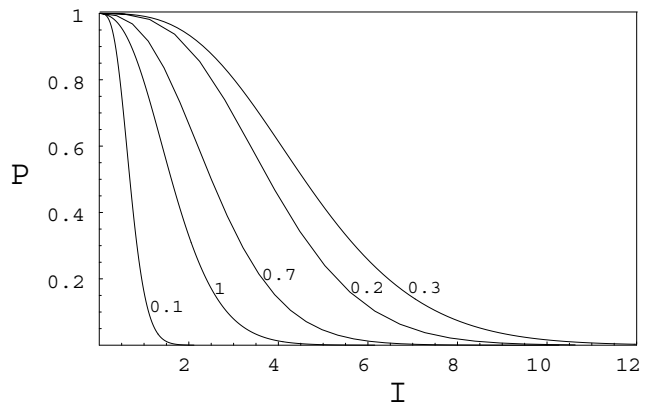}
\end{center}
\par
\vspace{-0.2cm} \caption{Survival probability $P$ versus the
number of stolen bits $I$. QDC
with repetition code $n=35$, and parameters $\Omega =1\simeq \tilde{\Omega}$%
, $c=1/2$ (so that $\bar{\protect\varepsilon}\simeq 1\%$ and $\mathcal{E}%
=1/35$). The curves refer to individual UGQCM attacks with different values
of added noise $\protect\sigma ^{2}$.}
\label{QCMcodePic}
\end{figure}

Now, let us consider the probability that Eve evades $M$ control modes.
Since the control mode is implemented exactly as before, we have again $\Pi
_{M}(\sigma ^{2})$ as in Eq.~(\ref{Survival_Probability}). Such a quantity
can be combined with the one of Eq.~(\ref{Eve_stolen_codes}). After $N$ runs
of the protocol, we have an average of $cN$ control modes and $(1-c)N$
message modes, so that Eve's survival probability is again $\Pi _{cN}(\sigma
^{2}):=P$ and the stolen information equal to $(1-c)NI_{AE}(\sigma ^{2}):=I$%
. Then, for every $\sigma ^{2}$, we can again evaluate the curve $P=P(I)$,
expressing Eve's survival probability as a function of the stolen bits.
According to Fig.~\ref{QCMcodePic}, the best choice for Eve is a UGQCM\ with
$\sigma ^{2}\simeq 0.3$, which enables her to steal about $10$ bits of
information before being detected. Such a result is a strong improvement
with respect to the basic protocol, where $80$ bits were left to Eve. Notice
that, for a low value of the noise like $\sigma ^{2}=0.1$, Eve gets $\simeq 1
$ bit while Alice transmits $\simeq 320$ bits of information by using $%
N\simeq 1.1\times 10^{4}$ systems. The maximal length of QDC is here bounded
by $4(1-c)(ncr)^{-1}\simeq 3500$ bits, i.e., $N\simeq 1.2\times 10^{5}$
quantum systems.

It is important to notice that the strong improvement brought by the
classical codes is proven provided the eavesdropping strategy is fixed,
i.e., Eve is restricted to an individual Gaussian attack where \textit{all}
the signal systems are attacked by a UGQCM. The idea of using repetition
codes is in fact based on the condition that \textit{all} the systems are
perturbed exactly in the same way. However, this is not true in general, and
we can design more appropriate strategies for Eve, which are specifically
optimized against the use of classical codes. This is the argument of the
following section.

\subsection{Non-Gaussian eavesdropping via intermittent attacks\label%
{IntermitSEC}}

In the previous Section~\ref{GaussFORcodes}, the QDC\ with repetition codes
has been tested against the same kind of attack considered for the basic
QDC. This attack is an individual UGQCM attack, which is indeed a Gaussian
attack if it is applied to every quantum system which is sent through the
channel. Notice that the same Gaussian interaction provided by the UGQCM
generates an overall non-Gaussian attack if it is applied to only a fraction
of the signal systems. This because an \textit{intermittent} use of Gaussian
interactions corresponds to the generation of an average non-Gaussian
interaction.

In this section, we introduce the notion of \textit{intermittent attacks}
which are individual non-Gaussian attacks based on the intermittent use of a
UGQCM. They are characterized by two parameters: the frequency parameter $%
\omega $ and the noise parameter $\sigma ^{2}$. The frequency parameter $%
\omega $ defines the probability that Eve attacks a signal system via a
UGQCM\ (and, then, detects the output clone via heterodyning). The noise
parameter $\sigma ^{2}$ defines the cloning noise variance which is
introduced by the UGQCM\ on the signals which are effectively attacked.
Then, for $N$ transmitted systems, a fraction $N\omega $\ is subject to
cloning interactions with noise $\sigma ^{2}$, while another fraction $%
N(1-\omega )$ is not perturbed by Eve. On average, Bob's output quadrature $%
x=q,p$ will follow the non-Gaussian distribution%
\begin{equation}
F_{\omega ,\sigma ^{2}}(x)=\omega G_{1+\sigma ^{2}}(x)+(1-\omega )G_{1}(x)~,
\end{equation}%
where $G_{\Delta }(x-\bar{x})$ is defined in Eq.~(\ref{Gaussian_Real}).
Clearly, in the particular case of $\omega =1$, this attack becomes Gaussian
and coincides with an individual UGQCM attack.

An intermittent attack can allow Eve to probe a subset of the systems very
heavily, instead of probing all the systems with a weaker interaction. This
peculiarity plays a non-trivial role in the case of QDC with repetition
codes, where the eavesdropping of a single bit of a codeword can be
sufficient to reconstruct all the encoded logical information. Here, we
explicitly show the superiority of the intermittent attacks against the use
of repetition codes. For the sake of simplicity, we consider only those
attacks whose frequencies can be written as $\omega =t/n$, where $n$ is the
length of the code and $t$ is an odd integer between $1$ and $n$.

After $N$ runs of the protocol, an intermittent attack of frequency $\omega $
(and noise $\sigma ^{2}$) will affect an average of $N\omega $ systems,
where $cN\omega $ are in CM and $(1-c)N\omega $ are in MM. Let us consider
the MM\ first. For each codeword of length $n$, there is an average of $%
t=n\omega $ bits attacked by Eve. Over these bits, Eve adopts the criterion
of majority voting in order to reconstruct the codeword. As a consequence,
the probability of a logical error is equal to the probability of having at
least $(t+1)/2$ bit flips, i.e.,%
\begin{equation}
P_{E}(t)=\sum_{k=\frac{t+1}{2}}^{t}\left(
\begin{array}{c}
t \\
k%
\end{array}%
\right) p^{k}(1-p)^{t-k}~,
\end{equation}%
where $p=p(\Delta _{E})$ is the single bit-flip probability in the Alice-Eve
channel, which is determined by $\Delta _{E}=1+(4\sigma ^{2})^{-1}$. Then,
for each MM, Eve extracts on average%
\begin{equation}
I_{AE}(\sigma ^{2},\omega )=2\{1-H[P_{E}(t)]\}/n
\end{equation}%
bits of information. After $N$ runs of the protocol, we have an average of $%
(1-c)N$ instances of MM and, therefore, Eve has stolen $I=(1-c)NI_{AE}(%
\sigma ^{2},\omega )$ bits of information. Now, let us consider the CM. For
each instance of CM, Bob performs two hypothesis tests, so that an average
of $2cN$ tests are done after $N$ runs of the protocol. Bob must distinguish
between the two hypotheses of Eq.~(\ref{Hp_Test}), which here means to
distinguish between the two distributions $F_{\omega ,0}(x)=G_{1}(x)$ (which
is Gaussian)\ and $F_{\omega ,\sigma ^{2}}(x)$ with $\sigma ^{2}\neq 0$
(which is non-Gaussian). Suppose that Bob knows exactly which are the
instances of CM that are attacked by Eve. This assumption clearly puts a
lower bound on the eavesdropping capabilities of Eve, which is however
sufficient to prove the result. In this case, Bob is able to isolate the $%
cN\omega $ attacked instances of CM from the $cN(1-\omega )$ instances which
are not attacked. On the attacked subset, Bob can now perform $2cN\omega $
tests in order to distinguish between two Gaussian distributions, i.e., $%
G_{1+\sigma ^{2}}(x)$ and $G_{1}(x)$. Then, we have to consider the
estimator of Eq.~(\ref{Estimator}), but now with $M=cN\omega $. As a
consequence, the survival probability of Eve after $N$ runs of the protocol
is now given by $P=\Pi _{cN\omega }(\sigma ^{2})$.

\begin{figure}[tbph]
\vspace{+0.3cm}
\par
\begin{center}
\includegraphics[width=0.47\textwidth] {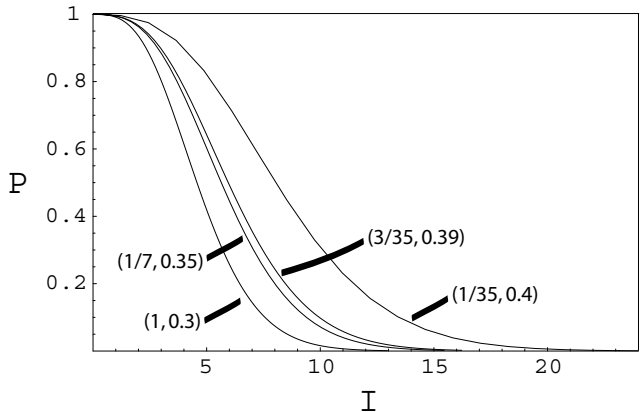}
\end{center}
\par
\vspace{-0.2cm} \caption{Survival probability $P$ versus the
number of stolen bits $I$. QDC
with repetition code $n=35$, and parameters $\Omega =1\simeq \tilde{\Omega}$%
, $c=1/2$ (so that $\bar{\protect\varepsilon}\simeq 1\%$ and $\mathcal{E}%
=1/35$). The curves refer to intermittent attacks with different parameters $%
(\protect\omega ,\protect\sigma ^{2})$. In particular, we have chosen $%
\protect\omega =1,1/7,3/35,1/35$ and the corresponding optimal noises $%
\protect\sigma ^{2}=0.3,0.35,0.39,0.4~.$}
\label{QCMcodeInterm}
\end{figure}

For every intermittent attack specified by the pair $\{\omega ,\sigma ^{2}\}$%
, we can now relate the survival probability $P$ to the number of stolen
bits $I$, i.e., we can consider the function $P=P(I)$. By adopting the
previous parameters for the QDC protocol, i.e., $n=35$, $\Omega =1$ and $%
c=1/2$, we derive the curves of Fig.~\ref{QCMcodeInterm}\ for different
values of the pair $\{\omega ,\sigma ^{2}\}$. In particular, we have chosen
the frequencies $\omega $ in the set $\{1,1/7,3/35,1/35\}$ and taken the
corresponding optimal noises $\sigma ^{2}$ which maximize Eve's stolen
information. As expected, the value of the optimal noise increases for
decreasing frequency. In particular, the best performance is achieved for $%
\omega =1/35$ (lowest frequency) and $\sigma ^{2}=0.4$ (highest noise),
where Eve is able to eavesdrop $20$ bits. Notice that this value is actually
a lower bound on Eve's capabilities, i.e., Eve is able to steal \textit{at
least} $20$ bits. In fact, except for the case $\omega =1$ (Gaussian
attack), all the curves are actually lower bounds on the actual performances
of Eve. Nevertheless, this is sufficient to prove the superiority of the
intermittent attacks in the eavesdropping of QDC with repetition codes.
Notice that the actual performances of these non-Gaussian attacks could be
much better. It is not excluded that they could completely annul the
advantages brought by the use of the classical codes.

\section{Conclusion\label{CONCLUsec}}

In this paper we have thoroughly reviewed the results of Ref.~\cite{QDCepl},
and we have also provided a deeper analysis of possible eavesdropping
strategies. In particular, the usage of classical correcting codes for QDC
leads to the birth of new kind of attacks, the intermittent attacks, which
are non-Gaussian and outperform the standard Gaussian attacks considered in
Ref.~\cite{QDCepl}. Because of this new strategy, the real advantages of
using classical codes for QDC are not completely clear. Despite this open
problem, the adoption of a basic QDC, with a suitable control mode
probability, always enables the honest users to decrease the number of
stolen bits to any desired value. Clearly, this is done at the expenses of
the efficiency of the protocol. This trade-off between the degree of privacy
and efficiency of the protocol is quite intuitive in our derivation. In
future work, it would be interesting to investigate the existence of a
precise relation between these two quantities. However, in order to derive
this kind of relation, the cryptoanalysis of the QDC should be first
extended to more general eavesdropping models, e.g., collective Gaussian
attacks involving the use of beam-splitters with thermal inputs, or more
general Gaussian interactions \cite{GaussInteractions}. At the present
stage, our protocols represent a simple proof-of-principle of a confidential
QDC in the framework of continuous variable systems, whose performances are
not definitive at all and could be greatly improved in future investigations.

As already discussed in Ref.~\cite{QDCepl}, our protocols for QDC allow an
effective communication only when a small amount of noise affects the
quantum channel, thus restricting their current application to relatively
short distances. Despite this restriction, there are however non-trivial
situations where they can be used in a profitable way. As explained in Ref.~%
\cite{QDCepl}, one of the possible applications is mutual entity
authentication \cite{Book}, where the two users identify each other by
comparing the bits of a pre-distributed and secret \textit{authentication key%
}. In this case, the usage of QDC is particularly profitable in the presence
of quantum impersonation attacks \cite{Dusek}, which are promptly revealed
by relatively small sessions of our basic protocol.

\section{Acknowledgements}

S.P. was supported by a Marie Curie Fellowship of the European
Community (Contract No. MOIF-CT-2006-039703). S.L. was supported
by the W.M. Keck foundation center for extreme quantum information
theory (xQIT).

\section{Appendix: possible variants for QDC\label{VariantsAPP}}

Here we briefly present several possible variants of the previous protocols.

\subsection{QDC using homodyne detector}

Simple variants of the previous protocols for QDC can be implemented via
homodyne instead of heterodyne detection. It is sufficient that Alice
encodes one single bit in the lattice by setting $U=U^{\prime }$. Then, Bob
randomly switches between $\hat{q}$ and $\hat{p}$ measurements, the exact
sequence being communicated to Alice at the end of the quantum
communication. In such a case, Eve is forced to a delayed-choice strategy,
where she has to keep all her ancillas before making the correct homodyne
measurement on each of them. Similar results can be easily proven for these
variants by considering that now the measurement noise is $\Delta =1/2$.

\subsection{Unifying control and message modes\label{CMeMMsection}}

Whenever the QDC is based on heterodyne detection and implemented with a
control mode probability $c=1/2$ (as in the case of the protocol of Sec.~\ref%
{sQDCcodedSection}), one can decide to distribute the control and message
modes on all the quantum systems. This is possible by randomly choosing a
quadrature for the encoding and the other for the check. Then, after Bob's
heterodyne detection, Alice declares the quadrature to be used for public
comparison.

\subsection{Postponed QDC}

In the basic protocol of Sec.~\ref{BasicSection}, run-by-run and after Bob's
detection, Alice declares which mode she has used (MM or CM) and the
corresponding classical information (mask amplitude $\alpha _{M}$ or signal
amplitude $\bar{\alpha}$). An alternative protocol consists in delaying this
declaration until the end of the quantum communication. At that point, Alice
will only declare the instances in CM and the corresponding amplitudes. Such
a procedure enables Alice and Bob to evaluate the noise of the channel
before revealing any confidential information. From such an estimation,
Alice computes the amount of information $I_{AE}$ that Eve can steal if she
unmasks the message. If $I_{AE}$ is negligible (according to a pre-agreed
tolerance level), then Alice unmasks all the message modes, communicating
her message to Bob. Otherwise, she has to abort. Alternatively, when $I_{AE}$
is not negligible but less than $I_{AB}$, Alice and Bob can possibly use the
remaining systems for distributing a secret key. Notice that such a
postponed protocol takes no advantage from the use of codes. Furthermore, it
can be simply implemented with $c=1/2$ and, therefore, also modified
according to Sec.~\ref{CMeMMsection}.


\begin{thebibliography}{99}
\bibitem{Exp} Notice that, in a realistic experimental setting, such an
intrinsic noise could include some \textquotedblleft
trusted\textquotedblright\ environmental noise which is not referable to Eve.
\bibitem{Integers} Ideally, the two integers $m$ and $m^{\prime }$ are both
uniformly distributed on $\mathbb{Z}$. More realistically, we can choose two
continuous values from a wide Gaussian distribution, and round them to the
nearest integers.
\bibitem{Bob} In order to be sure that Bob has received and detected the
state, Alice must require a classical communication from him.
\bibitem{Zero} A zero-tolerance protocol does not promptly stop in realistic
situations (where $\sigma ^{2}\neq 0$) because the underlying
hypothesis test is intrinsically imperfect (i.e., its probability
to fail is always non-zero).
\end{thebibliography}
\end{document}